\documentclass[10 pt,journal]{IEEEtran}
\usepackage{amsmath}
\usepackage{amssymb}
\usepackage{amsthm}
\usepackage{setspace}
\usepackage{pstricks}
\usepackage{psfrag}
\usepackage{graphicx}
\usepackage{caption}
\usepackage{subcaption}
\usepackage{booktabs}
\usepackage{cite}
\usepackage{booktabs}
\usepackage{multirow}
\usepackage{textcomp}
\usepackage{color}

\newcommand{\Define}{\stackrel{\Delta}{=}}

\newtheorem{mycorollary}{\bf Corollary}
\newtheorem{myproposition}{\bf Proposition}

\title{Uplink Performance of Time-Reversal MRC in Massive MIMO Systems Subject to Phase Noise}
\author{Antonios Pitarokoilis, Saif Khan Mohammed and Erik G. Larsson
\thanks{A. Pitarokoilis and Erik G. Larsson are with the Department of Electrical Engineering (ISY),
Link\"{o}ping University, 581 83 Link\"{o}ping, Sweden,
\texttt{\{antonispit,erik.larsson\}{@}isy.liu.se}. Saif K. Mohammed was with the Dept. of Electrical Engineering (ISY), Link\"{o}ping University, Sweden. He is now with the Dept. of Electrical Engineering, Indian Institute of Technology (I.I.T.) Delhi, India, \texttt{saifkm@ee.iitd.ac.in}.}
\thanks{This work was supported by the Swedish Foundation for Strategic Research (SSF) and ELLIIT. The work done by Saif K. Mohammed was supported by the Science and Engineering Research Board (SERB), Department of Science and Technology (DST), Government of India. This paper was presented in part at the 50th Allerton Conference on Communication, Control and Computing, Urbana-Champaign, IL, USA, Oct. 2012 \cite{Allerton12}.}}
\begin{document}

\maketitle
\begin{abstract}
Multi-user multiple-input multiple-output (MU-MIMO) cellular systems with an excess of base station (BS) antennas (Massive MIMO) offer unprecedented multiplexing gains and radiated energy efficiency. Oscillator phase noise is introduced in the transmitter and receiver radio frequency chains and severely degrades the performance of communication systems. We study the effect of oscillator phase noise in frequency-selective Massive MIMO systems with imperfect channel state information (CSI). In particular, we consider two distinct operation modes, namely when the phase noise processes at the $M$ BS antennas are identical (synchronous operation) and when they are independent (non-synchronous operation). We analyze a linear and low-complexity time-reversal maximum-ratio combining (TR-MRC) reception strategy. For both operation modes we derive a lower bound on the sum-capacity and we compare their performance. Based on the derived achievable sum-rates, we show that with the proposed receive processing an $O(\sqrt{M})$ array gain is achievable. Due to the phase noise drift the estimated effective channel becomes progressively outdated. Therefore, phase noise effectively limits the length of the interval used for data transmission and the number of scheduled users. The derived achievable rates provide insights into the optimum choice of the data interval length and the number of scheduled users.
\end{abstract}


\section{Introduction}\label{sec:intro}

Multiple-input multiple-output (MIMO) technology  offers substantial performance gains in  wireless links \cite{Foschini98}. The spatial degrees of freedom enable many users to share the same time-frequency resources, paving the way for multi-user MIMO (MU-MIMO) systems \cite{Gesbert07}. 
MU-MIMO systems with an excess of BS antennas, termed as  Massive MIMO or  large-scale MIMO, have recently attracted significant interest \cite{Marzetta10,Rusek13,CommMag13}. They promise a significant increase in the total cell throughput by means of simple signal processing. At the same time, the radiated power can be scaled down with the number of BS antennas, $M$, while maintaining a desired sum-rate. More specifically, in \cite{Hien12TComm} the authors show that in a MU-MIMO uplink with linear receivers and imperfect channel state information (CSI), by increasing the number of BS antennas from $1$ to $M$, one can reduce the total transmit power by a factor $O(\sqrt{M})$ while maintaining a fixed per-user information rate. In \cite{JacobJSAC13} the authors report an improved result for channels with arbitrary channel covariance matrices. The crucial assumption in Massive MIMO  is that the squared Euclidean norm of the channel vector of each user grows as $O(M)$, whereas the inner products between channel vectors of different users grow at a lesser rate. This assumption can be justified in the MU-MIMO setting since the users are typically separated by many wavelengths, which implies that their channel vectors become asymptotically (in the number of BS antennas) orthogonal. Extensive measurements have confirmed the validity of this assumption \cite{Rusek13,CommMag13}.

Phase noise is inevitable in communication systems due to imperfections in the circuitry of the local oscillators that are used for the conversion of the baseband signal to passband and vice versa. To be specific, phase noise is the instantaneous drift of the phase of the carrier wave and results in a widening of the power spectral density of the generated waveform. Phase noise causes a partial loss of coherency between the channel estimate and the true channel gain during data transmission. This can result in severe degradation of the system performance.

In MIMO an array power gain is obtained by coherently combining signals received by several antennas, using estimated channel responses. Since phase noise distorts the received data, it is crucial to examine its effect on the performance. Significant research work is available on phase noise. However, most of it is concerned with single-user single-antenna multi-carrier transmission, since multi-carrier transmission is more sensitive to phase noise compared to single-carrier transmission \cite{Pollet95}. In \cite{Tomba98} a method to calculate the bit-error-rate (BER) of a single-user orthogonal frequency division multiplexing (OFDM) system impaired with phase noise is provided. Reference \cite{WuBarNess} studies the signal-to-interference-and-noise-ratio (SINR) degradation in OFDM and proposes a method to mitigate the effect of phase noise. In \cite{Petrovic07} a method to characterize phase noise in OFDM systems is developed and an algorithm to compensate for the degradation is described. Finally, in \cite{Mehrpouyan12} the authors propose a method to jointly estimate the channel coefficients and the phase noise in a single-user MIMO system and an associated  phase noise mitigation algorithm.

From an information-theoretic point of view, the calculation of capacity of phase noise channels is challenging. To the best of our knowledge, the exact capacity of typical phase noise-impaired channels under realistic models is not known. The behavior of the capacity of such channels is only known asymptotically for some cases in the high signal-to-noise-ratio (SNR) regime \cite{Lapidoth}. In \cite{DurisiTComm14} the authors derive a non-asymptotic upper bound on the capacity of a single-user deterministic MIMO channel impaired with Wiener phase noise, which is tight in the high-SNR regime. In \cite{EmilHardware14}, the authors consider the performance of Massive MIMO systems with hardware impairments. Their model is suitable for the residual hardware impairments after the application of appropriate compensation algorithms. 

To the authors' knowledge, we present the first analysis of the effect of Wiener phase noise in a multi-user multi-antenna scenario with imperfect channel state information where single-carrier transmission is used. 
Specifically, we consider a single-cell frequency-selective MU-MIMO uplink, where a number of non-cooperative users transmit independent data streams to a base station having a large number of antennas. Since the channel is assumed to be unknown, CSI is acquired via uplink training. There are phase noise sources both at the transmitters and at the receiver. We consider and compare two distinct cases. In the first case, which is termed \emph{synchronous operation} mode, the phase noise processes at the BS antennas are identical. In the second case, which is termed \emph{non-synchronous operation} mode, the phase noise processes at the BS antennas are independent. These two operation modes correspond to the cases of a common phase reference versus independent phase references, respectively. A time-reversal maximum-ratio combining (TR-MRC) strategy is proposed and achievable sum-rates are derived for both operation modes.

Based on the derived expressions of the achievable sum-rates, we show that for a fixed desired per-user information rate, by doubling the number of BS antennas, the total transmit power can be reduced by a factor of $\sqrt{2}$. 
This is the same scaling law as without phase noise \cite{Hien12TComm}. We observe that the use of independent phase noise sources can yield higher sum-rate performance and we support this interesting result by a simple toy example for which the exact capacity is calculated. Furthermore, the achievable rate expressions reveal a fundamental trade-off between the length of the time interval spent on data transmission and the sum-rate performance. The rate expressions also provide valuable insight into the optimum number of scheduled users.

\section{System Model}\label{sec:sysmod}

We consider a frequency-selective MU-MIMO uplink channel with $M$ BS antennas and $K$ single-antenna users. The channel between the $k$-th user and the $m$-th BS antenna is modeled as a finite impulse response (FIR) filter with $L$ symbol-spaced channel taps. The $l$-th channel tap is given by $g_{m,k,l}\Define\sqrt{d_{k,l}}h_{m,k,l}$, where $h_{m,k,l}$ and $d_{k,l}$ model the fast and slow time-varying components, respectively. We assume a block fading model where $h_{m,k,l}$ is fixed during the transmission of a block of $N_c\Define N_D+(K+3)L-3$ symbols and varies independently from one block to another. $N_D$ denotes the number of channel uses utilized for data transmission (see Fig. \ref{fig:txsched}). We further assume that the channel fading process is ergodic. The parameters $d_{k,l}\geq 0,~l=0,\ldots,L-1$ model the power delay profile (PDP) of the frequency-selective channel for the $k$-th user. Since $\{d_{k,l}\}$ vary slowly with time and spatial location, we assume them to be fixed for the entire communication and independent of $m$. We further assume $h_{m,k,l}$ to be  independent and identically distributed (i.i.d.) zero-mean and unit-variance proper complex random variables. The i.i.d. assumption is justified in \cite{GaoVTC11,Rusek13,CommMag13}.\footnote{We note that with the i.i.d. assumption on the channel gains, the captured energy increases linearly with the number of BS antennas, $M$. This is not reasonable if $M$ grows unbounded. However, this deficiency of the model takes effect only for exorbitantly large values of $M$ which do not lie in the regime of our interest \cite{CommMag13},\cite{Rusek13},\cite{JacobJSAC13}.} Further, the PDP for every user is normalized such that the average received power is independent of the length of the channel impulse response, $L$. Therefore, it holds that
\vspace{-3mm}
\begin{align}\label{eq:PDPnorm}
 \sum_{l=0}^{L-1}\mathbb{E}\left[|\sqrt{d_{k,l}}h_{m,k,l}|^2\right]=\sum_{l=0}^{L-1}d_{k,l}=\alpha_k,
\end{align}
for $1\leq k\leq K$. The positive constants, $\alpha_k$, account for different propagation losses between users and are assumed to be fixed throughout the communication. The BS is assumed to have perfect knowledge of all the PDPs. Finally, we assume exact knowledge of the channel statistics at the BS, but not of the particular channel realizations.

\subsection{Phase Noise Model}\label{sub:PhNmodel}

Phase noise is introduced at the transmitter during up-conversion, when the baseband signal is multiplied with the carrier generated by the local oscillator. The phase of the generated carrier drifts randomly, resulting in a phase distortion of the transmitted signal. A similar phenomenon also happens at the receiver side during down-conversion of the bandpass signal to baseband. In the following, $\theta_k,~k=1,\ldots,K$ denotes the phase noise process at the $k$-th single-antenna user. Since the users have different local oscillators, the transmitter phase noise processes are assumed to be mutually independent. On the other hand, at the receiver side two distinct operation modes are considered. We term these operation modes as \emph{synchronous} and \emph{non-synchronous} operation depending on whether the phase noise processes at the BS antennas are identical or independent. For the synchronous case, all BS antennas are subject to the same phase noise process and $\phi$ denotes this common phase noise process at each BS antenna. This models the scenario of a centralized BS with a single oscillator feeding the down-conversion module in each receiver. For the case of non-synchronous operation, $\phi_m,~m=1,\ldots,M$ denotes the phase noise process at the $m$-th BS antenna. This models a completely distributed scenario where each BS antenna uses a distinct oscillator for down-conversion. We further assume that the phase noise processes $\theta_k,~k=1,\ldots,K$ and $\phi$ (or $\phi_m,~m=1,\ldots,M$) for the case of synchronous (or non-synchronous) operation mode are mutually independent.

In this study each phase noise process is modeled as an independent Wiener process, which is a 
well-established model \cite{Demir00, Petrovic07}. Therefore, the discrete-time phase noise process at the $k$-th user at time $i$ is given by\footnote{The discrete-time phase noise model is used since we will be working with the discrete-time complex baseband representation of the transmit and receive signals.}
\begin{align}\label{eq:phNWiener}
 \theta_k[i] = \theta_k[i-1] + w_k^t[i],
\end{align}
where $w_k^t[i]\sim\mathcal{N}(0,\sigma_\theta^2)$ are independent identically distributed zero-mean Gaussian increments with variance $\sigma_\theta^2\Define 4\pi^2f_c^2c_\theta T_s$, $f_c$ is the carrier frequency, $T_s$ is the symbol interval and $c_\theta$ is a constant that depends on the oscillator. Depending on the operation mode, the phase noise processes $\phi[i]$ and $\phi_m[i]$ at the $M$ BS antennas are defined in a manner similar to \eqref{eq:phNWiener}, where the increments have variance $\sigma_\phi^2\Define 4\pi^2f_c^2c_\phi T_s$.

\subsection{Received Signal}\label{sub:RecSig}

Let $x_k[i]$ be the symbol transmitted from the $k$-th user at time $i$. The received sample at the $m$-th BS antenna element at time $i$ is then given by, for the non-synchronous operation
\begin{align}\label{eq:RecSig_k2m_ns}
 y_m[i]=\sqrt{P}\sum_{k=1}^K\sum_{l=0}^{L-1}e^{-j\phi_m[i]}g_{m,k,l}e^{j\theta_k[i-l]}x_k[i-l]+n_m[i],
\end{align}
where $n_m[i]\sim\mathcal{CN}(0,\sigma^2)$ represents noise  at the $m$-th receiver at time $i$, which is distributed as circularly symmetric complex Gaussian.\footnote{In the following we will present only the expressions of the non-synchronous mode. The expressions for the synchronous operation are obtained easily by substituting $\phi_1[i]\equiv\ldots\equiv\phi_M[i]\equiv\phi[i]$. In Sections \ref{sec:AchSumRate}--\ref{sec:SeparatePhN}, when the expressions of the two distinct modes differ in a non-obvious way, both expressions will be given explicitly.} Each user transmits a stream of i.i.d. $\mathcal{CN}(0,1)$ information symbols (i.e., $x_k[i]\sim\mathcal{CN}(0,1)$), that are independent of the information symbols of the other users. $P$ denotes the average uplink transmitted power from each user.

\section{Transmission Scheme and Receive Processing}\label{sec:sched}

\begin{figure}
 \psfrag{K}[c][c]{\scriptsize$KL$}
 \psfrag{L}[c][c]{\scriptsize$L-1$}
 \psfrag{N}[c][c]{\scriptsize$N_D$}
 \psfrag{Training}[c][c]{\scriptsize Training}
 \psfrag{Data}[c][c]{\scriptsize Data phase}
 \psfrag{Preamble}[c][c]{\scriptsize Preamble}
 \psfrag{Postamble}[c][c]{\scriptsize Postamble}
 \psfrag{IBI}[c][c]{\scriptsize IBI}
 \centering
 \includegraphics[width=0.5\textwidth]{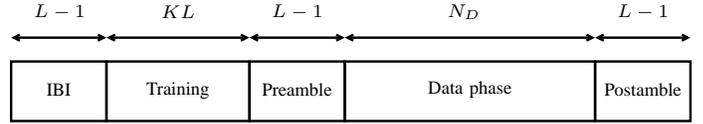}
\caption{The transmission block is assumed to span a coherence interval, $N_c\Define N_D+(K+3)L-3$. In each block, the first $KL$ channel uses (cu) are utilized for pilot based channel estimation and $N_D$ cu are utilized for data transmission. An all-zero block, a preamble and a postamble of $L-1$ cu each are added due to the edge effects of the channel.}\label{fig:txsched}
\end{figure}

We consider a block-based uplink transmission scheme. A transmission block of $N_c$ channel uses consists of $KL$ channel uses dedicated to uplink channel training followed by a preamble of  $L-1$ channel uses, where i.i.d. $\mathcal{CN}(0,1)$ non-information symbols are sent. The data interval of $N_D$ channel uses comes after that and a postamble of $L-1$ channel uses is appended at the end of the coherence interval, where i.i.d. $\mathcal{CN}(0,1)$ non-information symbols are sent. The inclusion of the preamble and postamble accounts for the edge effects introduced due to the intersymbol interference. This way the subsequent analysis is valid for all the $N_D$ channel uses during data transmission and no separate analysis for the edges of the data interval is required. At the beginning of each coherence interval an all-zero block of $L-1$ channel uses is prepended to eliminate inter-block interference (IBI) (see Fig. \ref{fig:txsched}).

\subsection{Channel Estimation}\label{sub:ChanEst}

For coherent demodulation, the BS needs to estimate the uplink channel. This is facilitated through the transmission of uplink pilot symbols during the training phase of each transmission block.\footnote{In this paper we deal only with uplink transmission. In Massive MIMO Time Division Duplex (TDD) operation pilots are transmitted on the uplink. The number of required pilots scales with the number of terminals, $K$, but not the number of BS antennas, $M$, making Massive MIMO scalable with respect to $M$ \cite{Marzetta10},\cite{Rusek13}.} The users transmit uplink training signals sequentially in time, i.e., at any given time only one user is transmitting uplink training signals and all other users are silent. To be precise, the $k$-th user sends an impulse of amplitude $\sqrt{P_p KL}$ at the $(k-1)L$-th channel use and is idle for the remaining portion of the training phase. Here, $P_p$ is the average power transmitted by a user during the training phase. We choose the proposed training sequence since it allows for a very simple channel estimation scheme at the BS and since it facilitates our derivation of achievable rates. However, many of our results, such as partial loss of coherency due to Wiener phase noise and monotonic decrease in performance with increased variance of the phase noise increments, are expected to be qualitatively valid also for other (but not necessarily all possible) training schemes. Therefore, using \eqref{eq:RecSig_k2m_ns}, the signal received at the $m$-th  BS receiver at time $(k-1)L+l,~l=0,\ldots,L-1,~k=1,\ldots,K$ is given by, for non-synchronous operation
\begin{align}\label{eq:recSigChanEstNS}
 y_m[(k-1)L+l]\!&=\!\sqrt{P_p KL}g_{m,k,l}e^{j(\theta_k[(k-1)L]-\phi_m[(k-1)L+l])}\nonumber\\
 &+n_m[(k-1)L+l].
\end{align}
\normalsize
Based on \eqref{eq:recSigChanEstNS}, we derive the maximum likelihood (ML) estimate of the effective channel $g_{m,k,l}e^{j(\theta_k[(k-1)L]-\phi_m[(k-1)L+l])}$. The corresponding channel estimates are then given by, for non-synchronous operation
\begin{align}\label{eq:chanEstNS}
 \hat g_{m,k,l}&=\frac{1}{\sqrt{P_p KL}}y_m[(k-1)L+l]\nonumber\\&=g_{m,k,l}e^{-j\phi_m[(k-1)L+l]}e^{j\theta_k[(k-1)L]}\nonumber\\
 &+\frac{1}{\sqrt{P_p KL}}n_m[(k-1)L+l].
\end{align}
We observe that the channel estimate is distorted by the AWGN and by the phase noise of the local oscillators at the user and at the BS.

\subsection{Time-Reversal Maximum Ratio Combining (TR-MRC)}\label{sub:MRC}

Using \eqref{eq:RecSig_k2m_ns}, the received signal during the data phase is given by, for non-synchronous operation
\begin{align}\label{eq:RecSigDataNS}
 y_m[i]=\sqrt{P_D}\sum_{k=1}^K\sum_{l=0}^{L-1}e^{-j\phi_m[i]}g_{m,k,l}e^{j\theta_k[i-l]}x_k[i-l]+n_m[i],
\end{align}
where $i\in\mathcal{I}_d,~\mathcal{I}_d\Define\{(K+1)L-1,\ldots,(K+1)L+N_D-2\}$ and $P_D$ is the per-user average transmit power constraint during the data phase. Motivated by the need for low-complexity detection, we consider the TR-MRC receiver at the BS. The TR-MRC receiver convolves  the received symbols, $y_m[i]$,   with the complex conjugate of the time-reversed estimated channel impulse response. The detected symbol, $\hat x_k[i]$, is given by
\begin{align}\label{eq:detectSym}
 \hat x_k[i]&=\sum_{l=0}^{L-1}\sum_{m=1}^M \hat g^*_{m,k,l}y_m[i+l],
\end{align}
where $(\cdot)^*$ denotes the complex conjugation operation.
\section{Achievable sum-rate}\label{sec:AchSumRate}

We use the information sum-rate as the performance metric for quantifying the effects of phase noise. To this end, using \eqref{eq:chanEstNS} and \eqref{eq:RecSigDataNS} for the non-synchronous operation, \eqref{eq:detectSym} is written as
\begin{align}\label{eq:sysmodcomponents}
 \hat x_k[i] &= A_k[i]x_k[i]+\texttt{ISI}_k[i]+\texttt{MUI}_k[i]+\texttt{AN}_k[i],
\end{align}
where it holds for the non-synchronous operation that
\begin{align}\label{eq:des_sig_coef}
 A_k[i]&\Define\sqrt{P_D}\sum_{m=1}^{M}\sum_{l=0}^{L-1}\left|g_{m,k,l}\right|^2\vartheta\!\left(\!\substack{m,k,k\\i,l,l}\!\right)
\end{align}
\begin{align}\label{eq:ISI_term}
 \texttt{ISI}_k[i]\!\Define\!\sqrt{P_D}\!\sum_{m=1}^M\!\sum_{l=0}^{L-1}\!\sum_{\substack{p=0\\p\neq l}}^{L-1}\!g^*_{m,k,l}g_{m,k,p}\vartheta\!\left(\!\substack{m,k,k\\i,l,p}\!\right)\!x_k[i\!+\!l\!-\!p]
\end{align}
\begin{align}\label{eq:MUI_term}
 \texttt{MUI}_k[i]&\Define\sqrt{P_D}\sum_{m=1}^{M}\sum_{\substack{q=1\\ q\neq k}}^K\sum_{l=0}^{L-1}\sum_{p=0}^{L-1}g^*_{m,k,l}g_{m,q,p}\times\nonumber\\
 &\vartheta\!\left(\!\substack{m,k,q\\i,l,p}\!\right)x_q[i+l-p]
\end{align}
\begin{align}\label{eq:AN_term}
 \texttt{AN}_k[i]&\Define\sqrt{\frac{P_D}{P_pKL}}\sum_{m=1}^M\sum_{q=1}^K\sum_{l=0}^{L-1}\sum_{p=0}^{L-1}g_{m,q,p}\times\nonumber\\& e^{-j(\phi_m[i+l]-\theta_q[i+l-p])}n_m[(k-1)L+l]x_q[i\!+\!l\!-\!p]\nonumber\\
 &+\sum_{m=1}^M\sum_{l=0}^{L-1}\hat g^*_{m,k,l}n_m[i+l],
\end{align}
where $\vartheta\!\left(\!\substack{m,k,q\\i,l,p}\!\right)\!\!\Define\!\! e^{j(\theta_q[i+l-p]-\theta_k[(k-1)L]-\phi_m[i+l]+\phi_m[(k-1)L+l])}$. In \eqref{eq:sysmodcomponents}, $A_k[i]x_k[i]$ is the desired signal term for the $k$-th user, $\texttt{ISI}_k[i]$ stands for the intersymbol interference for user $k$ at time $i$, caused by the information symbols of the $k$-th user transmitted at other time instances, $\texttt{MUI}_k[i]$ denotes the multi-user interference due to the information symbols of the other users and finally $\texttt{AN}_k[i]$ is an aggregate noise term that incorporates the effects of the channel estimation error and the receiver AWGN noise, $n_m[i]$. The expressions for the terms in \eqref{eq:sysmodcomponents} for the synchronous operation are obtained from \eqref{eq:des_sig_coef}-\eqref{eq:AN_term} by substituting $\phi_1[i]\equiv\ldots\equiv\phi_M[i]\equiv\phi[i]$.

In the following, we derive an achievable information rate for the $k$-th user. Similar capacity bounding techniques have been used earlier in e.g. \cite{Hassibi03, Marzetta06}. In \eqref{eq:sysmodcomponents}, we add and subtract the term $\mathbb{E}\left[A_k[i]\right]x_k[i]$, where the expectation is taken over the channel gains, $g_{m,k,l}$, and the phase noise processes, $\theta_k,~\phi$ for the synchronous operation and $\theta_k,~\phi_m$ for the non-synchronous operation. We relegate the variation around this term, i.e., $\texttt{IF}_k[i]\Define(A_k[i]-\mathbb{E}\left[A_k[i]\right])x_k[i]$, to an effective noise term. This results in the following equivalent expression
\begin{align}\label{eq:rateSysMod}
 \hat x_k[i] &= \mathbb{E}\left[A_k[i]\right]x_k[i]+\texttt{EN}_k[i],
\end{align}
where 
\begin{align}\label{eq:effective_noise}
\texttt{EN}_k[i]\Define \texttt{IF}_k[i]+\texttt{ISI}_k[i]+\texttt{MUI}_k[i]+\texttt{AN}_k[i],
\end{align}
is the effective additive noise term. In  \eqref{eq:rateSysMod} the detected symbol, $\hat x_k[i]$, is a sum of two \emph{uncorrelated} terms (i.e., $\mathbb{E}\left[\left(\mathbb{E}[A_k[i]]x_k[i]\right)\left(\texttt{EN}_k[i]\right)^*\right] = 0$). The importance of the equivalent representation in \eqref{eq:rateSysMod} is that the scaling factor $\mathbb{E}[A_k[i]]x_k[i]$ of the desired information symbol is a constant, which is known at the BS since the BS has knowledge of the channel statistics. The exact probability distribution of $\texttt{EN}_k[i]$ is difficult to compute. However, its variance can be easily calculated given that the channel statistics is known at the BS. Therefore,  \eqref{eq:rateSysMod} describes an effective single-user single-input single-output (SISO) additive noise channel, where the noise is zero mean, has 
known variance and is uncorrelated with the desired signal term. From the expressions for $A_k[i]$ and $\texttt{EN}_k[i]$ in \eqref{eq:des_sig_coef} and \eqref{eq:effective_noise}, the mean value of $A_k[i]$ and the variance of $\texttt{EN}_k[i]$ is given by two propositions that follow.
\begin{myproposition}\label{prop:meanExpressions}
The mean value of $A_k[i]$ in both operation modes is given by
\begin{align}\label{eq:meanAs}
 \mathbb{E}[A_k[i]] = \sqrt{P_D}M\alpha_k e^{-\frac{\sigma_\phi^2+\sigma_\theta^2}{2}(i-(k-1)L)}.
\end{align}
\end{myproposition}
\begin{IEEEproof}
We prove the statement for the non-synchronous operation. The proof for the synchronous operation is nearly identical. From \eqref{eq:des_sig_coef}, we have
\begin{align*}
 \mathbb{E}&[A_k[i]]=\mathbb{E}\left[\sqrt{P_D}\sum_{m=1}^M\sum_{l=0}^{L-1}\left|g_{m,k,l}\right|^2\vartheta\!\left(\!\substack{m,k,k\\i,l,l}\!\right)\right]\\
 &\stackrel{(a)}{=}\sqrt{P_D}\mathbb{E}\left[e^{-j\left(\theta_k[(k-1)L]-\theta_k[i]\right)}\right]\sum_{m=1}^M\sum_{l=0}^{L-1}\mathbb{E}\left[\left|g_{m,k,l}\right|^2\right]\nonumber\\
 &\cdot\mathbb{E}\left[e^{-j\left(\phi_m[i+l]-\phi_m[(k-1)L+l]\right)}\right]\nonumber\\
 &\stackrel{(b)}{=}\sqrt{P_D}e^{-\frac{\sigma_\theta^2}{2}(i-(k-1)L)}\sum_{m=1}^M\sum_{l=0}^{L-1}d_{k,l}e^{-\frac{\sigma_\phi^2}{2}(i-(k-1)L)} \nonumber\\
 &\stackrel{(c)}{=}\sqrt{P_D}M\alpha_k e^{-\frac{\sigma_\phi^2 + \sigma_\theta^2}{2}(i-(k-1)L)}.
\end{align*}
In (a) we have used the fact that the channel realizations, $g_{m,k,l}$, the phase noise at the BS, $\phi_m$, and the phase noise at the $k$-th user, $\theta_k$, are mutually independent random processes. The equality (b) is a consequence of the Wiener phase noise model. That is, after a time interval, $\Delta t=i-(k-1)L$, the phase drift of an oscillator is a zero mean Gaussian random variable with variance that is proportional to $\Delta t$,
\begin{align*}
 U_{\phi_m}&\Define \phi_m[i\!+\!l]\!-\!\phi_m[(k\!-\!1)L\!+\!l]\!\sim\!\mathcal{N}(0,\sigma_\phi^2(i-(k-1)L)),\\
 U_{\theta_k}&\Define \theta_k[i]-\theta_k[(k-1)L]\sim\mathcal{N}(0,\sigma_\theta^2(i-(k-1)L)).
\end{align*}
Henceforth $\mathbb{E}\left[e^{-j U_{\phi_m}}\right]=\varphi_{\phi_m}(-1)=e^{-\frac{\sigma_\phi^2}{2}(i-(k-1)L)}$ and $\mathbb{E}\left[e^{jU_{\theta_k}}\right]=\varphi_{\theta_k}(1)=e^{-\frac{\sigma_\theta^2}{2}(i-(k-1)L)}$, where $\varphi_{\phi_m}(\cdot)$ and $\varphi_{\theta_k}(\cdot)$ are the characteristic functions of $U_{\phi_m}$ and $U_{\theta_k}$, respectively. The equality (c) follows from \eqref{eq:PDPnorm}.
\end{IEEEproof}

In \eqref{eq:meanAs}, the factor $M$ signifies the combining gain in a coherent receiver (i.e.,  when $\sigma_\phi=\sigma_\theta=0)$. The factor $e^{-\frac{\sigma_\phi^2+\sigma_\theta^2}{2}(i-(k-1)L)}$ signifies the loss in effective amplitude gain due to the non-coherency between the received data samples and the estimated channel gains. Note that this non-coherency arises due to the fact that the channel gains for the $k$-th user are estimated at $t=(k-1)L+l,~l=0,\ldots,L-1$ and the samples for detecting $x_k[i]$ are received at $t=i+l,~l=0,\ldots,L-1$, that is,
 $i-(k-1)L$ samples later. The oscillator phase drift in this time period results in a partial non-coherency. It is clear that the larger this time difference is the smaller the effective amplitude gain is (the effective amplitude $M\alpha_k e^{-\frac{\sigma_\phi^2+\sigma_\theta^2}{2}(i-(k-1)L)}$ decreases exponentially with increasing time difference $i-(k-1)L$).
 
\begin{myproposition}\label{prop:varExpressions}
The variance $\texttt{Var}(\texttt{EN}_k[i])\Define\mathbb{E}\left[|\texttt{EN}_k[i]-\mathbb{E}\left[\texttt{EN}_k[i]\right]|^2\right]$ satisfies, for synchronous operation
\begin{align}\label{eq:VarENs}
 \varsigma_k^s[i]\Define\texttt{Var}(\texttt{EN}^s_k[i])&= P_DM^2\kappa_k[i]+C_k,
\end{align}
and for non-synchronous operation
\begin{align}\label{eq:VarENns}
  \varsigma_k^{ns}[i]\Define\texttt{Var}(\texttt{EN}^{ns}_k[i])&= P_DM^2\alpha_k^2\varpi_k[i]+P_DM\xi_k[i]+C_k,
\end{align}
where $\kappa_k[i]\Define\sum_{l=0}^{L-1}\sum_{l'=0}^{L-1}d_{k,l}d_{k,l'}e^{-\sigma_\phi^2|l-l'|}-\alpha_k^2e^{-(\sigma_\phi^2+\sigma_\theta^2)(i-(k-1)L)}$,\\ $\xi_k[i]\Define\sum_{l=0}^{L-1}\sum_{l'=0}^{L-1}d_{k,l}d_{k,l'}e^{-\sigma_\phi^2|l-l'|}-\alpha_k^2e^{-\sigma_\phi^2(i-(k-1)L)}$, $\varpi_k[i]\Define e^{-\sigma_\phi^2(i-(k-1)L)}\left(1-e^{-\sigma_\theta^2(i-(k-1)L)}\right)$,\\ \small{$C_k\Define P_DM\alpha_k\sum_{q=1}^{K}\alpha_q+\sigma^2 M\left(\frac{P_D}{P_p K}\sum_{q=1}^{K}\alpha_q+\alpha_k+\frac{\sigma^2}{K P_p}\right)$.}
\end{myproposition}
\normalsize
\begin{IEEEproof}
See the Appendix.
\end{IEEEproof}
The second term of the constant $C_k$ in Proposition \ref{prop:varExpressions} is the contribution of the additive noise term $\texttt{AN}_k[i]$.   This contribution has variance $\mathbb{E}\left[\left|\texttt{AN}_k[i]\right|^2\right]=\sigma^2 M\left(\frac{P_D}{P_p K}\sum_{q=1}^{K}\alpha_q+\alpha_k+\frac{\sigma^2}{K P_p}\right)$.
The term $\sigma^2M\frac{P_D}{P_p K}\sum_{q=1}^{K}\alpha_q$ corresponds to the cross-correlation between the channel estimation error  in \eqref{eq:chanEstNS} and the received symbols in \eqref{eq:RecSigDataNS}. 
 The term $\sigma^2M\alpha_k$ corresponds to the filtered noise  \eqref{eq:detectSym}. 
 Finally, the last term $\sigma^2M\frac{\sigma^2}{KP_p}$ corresponds to the variance of the channel estimation error.

In the following we provide a coding strategy that justifies the achievable rates we are interested in deriving. From Propositions \ref{prop:meanExpressions} and \ref{prop:varExpressions}, it is obvious that $\mathbb{E}[A_k[i]]$ and $\texttt{Var}(\texttt{EN}_k[i])$ depend on $i$ and are different for different $i\in\mathcal{I}_d$. Further, for a given $i$, across multiple transmission blocks, the terms $\mathbb{E}[A_k[i]]$ and $\texttt{Var}(\texttt{EN}_k[i])$ are the same and the realizations of $\texttt{EN}_k[i]$ are i.i.d. Hence, for each $i$, we have an additive noise SISO channel. This motivates us to consider $N_D$ channel codes for each user, one for each $i\in\mathcal{I}_d$. At the $k$-th transmitter (user), the symbols of the $i$-th channel code ($x_k[i]$) are transmitted only during the $i$-th channel use of each transmission block. Similarly, at the BS, for the $k$-th user, the $i$-th received and processed symbols (i.e., $\hat x_k[i]$) across different transmission blocks are jointly decoded. Essentially, this implies that, at the BS we have $N_D$ parallel channel decoders for each user. We propose the above scheme of $N_D$ parallel channel codes for each user only to derive a lower bound on the achievable information rate. In practice, due to reasons of complexity, channel coding/decoding would not only be performed across different transmission blocks, but also across consecutive channel uses within each transmission block.

Given the previously described coding strategy, we are now interested in computing a lower bound on the reliable rate of communication for each of the $N_D$ channel codes. Since the data symbols $x_k[i]$ are Gaussian, for each $i\in\mathcal{I}_d$ a lower bound on the information rate for the effective channel in \eqref{eq:rateSysMod} can be computed by considering the worst case (in terms of mutual information) uncorrelated additive noise. With Gaussian information symbols, it is known that the worst case uncorrelated noise is Gaussian with the same variance as that of $\texttt{EN}_k[i]$ \cite{Hassibi03}. Consequently, a lower bound on $I(\hat x_k[i];x_k[i])$ (i.e., the mutual information rate for the $i$-th channel code for user $k$) is given by Proposition \ref{prop:AchRates}.
\begin{myproposition}\label{prop:AchRates}
 The achievable rate for the $i$-th channel code for the $k$-th user is given by
\begin{align}\label{eq:achRate}
 I(\hat x_k[i]&;x_k[i])\geq R_k^\times[i]\nonumber\\
 &\Define\log_2\left(1+\frac{P_DM^2\alpha_k^2 e^{-(\sigma_\phi^2+\sigma_\theta^2)(i-(k-1)L)}}{\varsigma_k^\times[i]}\right),
\end{align}
where $\times=s$ for synchronous operation and $\times=ns$ for non-synchronous operation and $\varsigma_k^\times$ are given in Proposition \ref{prop:varExpressions}.
\end{myproposition}

\begin{mycorollary}\label{cr:nsynch_is_better}
 Based on the   lower bounds \eqref{eq:achRate}, the proposed TR-MRC receiver exhibits better performance in the case of non-synchronous operation.
\end{mycorollary}
\begin{IEEEproof}[Proof of Corollary \ref{cr:nsynch_is_better}]
\begin{align*}
\varsigma_k^s&[i]-\varsigma_k^{ns}[i]=P_DM(M-1)\xi_k[i]\\
&\stackrel{(a)}{\geq} P_DM(M-1)\left(\alpha_k^2 e^{-\sigma_\phi^2 (L-1)}-\alpha_k^2e^{-\sigma_\phi^2(i-(k-1)L)}\right)\\
&\stackrel{(b)}{\geq} P_DM(M-1)\left(\alpha_k^2 e^{-\sigma_\phi^2 (L-1)}-\alpha_k^2e^{-\sigma_\phi^2(2L-1)}\right)\geq 0.
\end{align*}
The inequality $(a)$ follows from the fact that $|l-l'|\leq L-1$ and \eqref{eq:PDPnorm}. The inequality $(b)$ follows
since  $i\geq KL+L-1\Rightarrow i-(k-1)L\geq (K-k)L + 2L-1\geq 2L-1$ and $k\leq K$.
\end{IEEEproof}

Note that Corollary~\ref{cr:nsynch_is_better} compares two lower bounds. However, there are good reasons to expect that these lower bounds are actually quite good predictions of the performance that could be
achieved in reality. This is so because substantially we make a Gaussianity assumption on the effective noise. This is also very likely
the type of approximation that would be used when deriving a soft decoding (LLR) metric for insertion into for example, a turbo decoder.
Hence, using this Gaussian approximation would predict quite well the performance achievable with good channel codes and standard
decoding metrics assuming Gaussian noise. Also note that comparing lower bounds that are reasonably tight is a standard practice in the communication theory literature. 

\begin{figure}
 \psfrag{X}{$X$}
 \psfrag{t1}{$e^{j\varphi_1}$}
 \psfrag{t2}{$e^{j\varphi_2}$}
 \psfrag{y1}{$Y_1$}
 \psfrag{y2}{$Y_2$}
 \psfrag{Y}{$Y$}
 \psfrag{half}{$1/2$}
 \centering
 \includegraphics[height=0.15\textheight]{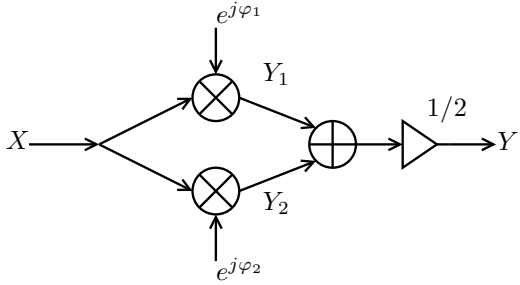}
 \caption{System model for the example.}\label{fig:toyExample}
\end{figure}
Corollary \ref{cr:nsynch_is_better} conveys an interesting result that the performance is better when the phase noise processes at the different BS antennas are uncorrelated. However, this is not the first time that such a result is reported. In \cite[Section III.A]{Hoehne10} the authors study the effect of phase noise in single-user beamforming. The performance measure they use is the error vector magnitude (EVM) and they show that EVM is smallest in the desired direction when uncorrelated phase noise sources are used. In \cite[Section VI.D]{EmilHardware14} the authors consider the impact of phase noise distortion in a flat fading channel with maximum ratio combining, using a small phase noise approximation. They also observe that by using separate oscillators the distortion scales as $O(t)$, where $t$ is the time   elapsed from channel estimation to data detection. On the other hand, when a common oscillator is used the distortion scales as $O(tM)$. 
(Note that in contrast to our analysis, \cite{EmilHardware14} used a much simpler model that did not include the effects of intersymbol interference, nor of multiuser interference.)
From Corollary \ref{cr:nsynch_is_better}, it can be argued that the use of independent oscillators at the BS can be beneficial when TR-MRC is used. Also, for a desired sum-rate performance one can choose between a high quality single oscillator or many oscillators of lower quality.

\subsubsection{Achievable Sum-Rate}

Since no data transmission happens during the training phase, the overall effective information rate achievable by the $k$-th user is given by,  
\begin{align}\label{eq:userRateNS}
R_k^{ \times}\Define\frac{1}{N_c}\sum_{i\in\mathcal{I}_d}R_k^{ \times}[i].
\end{align}
The achievable sum-rate is therefore given by 
\begin{align}\label{eq:sumRateNS}
 R^{ \times} = \sum_{k=1}^K R_k^{ \times}=\frac{1}{N_c}\sum_{k=1}^K\sum_{i\in\mathcal{I}_d}R_k^{ \times}[i].
\end{align}

It is clear that  phase noise degrades the sum-rate performance both with
synchronous and non-synchronous operation. To see this formally, note that
the sum-rate for the no-phase-noise case can be derived from \eqref{eq:achRate}, \eqref{eq:userRateNS} and \eqref{eq:sumRateNS} by setting $\sigma_\phi^2=\sigma_\theta^2=0$ and is given by 
\begin{align}\label{eq:sumRateNophN}
 \mathcal{R}&=\frac{N_D}{N_c}\sum_{k=1}^{K}\log_2\left(1+\frac{P_DM^2\alpha_k^2}{C_k}\right).
\end{align}
Since, $\frac{P_D}{\sigma^2}M^2\alpha_k^2\geq \frac{P_D}{\sigma^2}M^2\alpha_k^2 e^{-(\sigma_\phi^2+\sigma_\theta^2)(i-(k-1)L)}$ and $\varsigma_k^s[i]\geq \varsigma_k^{ns}[i]\geq C_k$ we have that $\mathcal{R}\geq R^\times$.

\subsection{Exact Analysis of Synchronous versus Non-Synchronous Operation for a Toy Channel Model}

In the following, we provide a simple example to illustrate that the
conclusion drawn from Corollary \ref{cr:nsynch_is_better} is the result of a fundamental phenomenon and not an artifact of the   techniques used to derive the lower bounds on the information rate.
We consider a very simple channel with only phase noise and no AWGN, see Fig. \ref{fig:toyExample}. Here $X\in\{\pm 1\},~\Pr\{X=+1\}=p,~\Pr\{X=-1\}=1-p$ is the input to the channel. The input $X$ is rotated by $\varphi_1$ and $\varphi_2$ to form $Y_1$ and $Y_2$, respectively. Let the random variables $\varphi_1,~\varphi_2$  model the  phase noise, with the following probability mass functions (p.m.f.): $\varphi_i\in\{-\frac{\pi}{2},0,\frac{\pi}{2}\},~\Pr\{\varphi_i=-\frac{\pi}{2}\}=\Pr\{\varphi_i=0\}=\Pr\{\varphi_i=\frac{\pi}{2}\}=\frac{1}{3},~i=1,2$. The output of this discrete memoryless channel (DMC) is given by
\begin{align}\label{eq:toySysMod}
 Y=\frac{1}{2}\left(e^{j\varphi_1}+e^{j\varphi_2}\right)X.
\end{align}

We now consider two cases, firstly when the two phase noise processes are synchronous (i.e., $\varphi_1\equiv\varphi_2$) and secondly when they are non-synchronous and mutually independent. In the  synchronous case, $\varphi_1\equiv \varphi_2$ so $Y=e^{j\varphi_1}X$. Then $Y$ takes values in $\mathcal{Y}_s=\{+1,+j,-1,-j\}$. The  output symbols have the p.m.f.: $\Pr\{Y=+1\}=p/3$, $\Pr\{Y=-1\}=(1-p)/3$, $\Pr\{Y=\pm j\}=1/3$. The capacity of this channel can be calculated as follows

\begin{align*} 
 C_s&=\max_p I(X;Y)=\max_p H(Y)-H(Y|X)\\&=\max_p\frac{1}{3}H_2(p)=1/3\text{ bits,}
\end{align*}
where $H_2(p)$ is the binary entropy function.  

In the  non-synchronous case, where $\varphi_1$ and $\varphi_2$ are independent of each other, the output variable takes values in $\mathcal{Y}_{ns}=\{+1,\frac{1}{2}(1+j),\frac{1}{2}(1-j),j,0,-j,-\frac{1}{2}(1-j),-\frac{1}{2}(1+j),-1\}$. The p.m.f. of the output  is $\Pr\{Y=+1\}=p/9$, $\Pr\{Y=(1\pm j)/2\}=2p/9$, $\Pr\{Y=\pm j\}=1/9$, $\Pr\{Y=0\}=2/9$, $\Pr\{Y=-(1\pm j)/2\}=2(1-p)/9$, and $\Pr\{Y=-1\}=(1-p)/9$. We find that $H(Y)=\frac{5}{9}H_2(p)+\log_2 9 -6/9$ and $H(Y|X=\pm 1)=\log_2 9 -6/9$. Then, the capacity is given by

\begin{align*} 
C_{ns}&=\max_p I(X;Y)=\max_p H(Y)-H(Y|X)\\&=\max_p\frac{5}{9}H_2(p)=5/9\text{ bits.}
\end{align*}
Since $C_s<C_{ns}$, it is concluded that the capacity of the channel in Fig. \ref{fig:toyExample} 
is strictly larger in the non-synchronous case than in the synchronous case.

Note that the example does not show that the capacity \emph{always} increases if we use independent phase noise sources. However, it shows  that \emph{there are} cases where the use of independent phase noise sources can be beneficial.

\section{Asymptotic Results}\label{sec:mainResults}

The achievable rates presented in Proposition \ref{prop:AchRates} hold for any $M$. In this section we present some asymptotic (in $M$) results based on these achievable rates in order to investigate the Massive MIMO effect in the system under study. In the following $\beta\Define\frac{P_p}{P_D}>0$ denotes the ratio between the per-user average transmit power during the training phase and   during the transmission phase.

We first note that  in the low SNR regime, the performance loss due to phase noise is not significant.
To see this quantitatively, consider  the sum-rate  when phase noise is present,  given by \eqref{eq:sumRateNS}. From \eqref{eq:achRate} it is clear that in the low-SNR regime, i.e., when $P_D/\sigma^2\ll 1$, the dominating factor in the denominator of the argument of the $\log_2$ function is, in both operation modes, the term $\frac{\sigma^4M}{K \beta P_D}$. From \eqref{eq:sumRateNophN} (after the substitution $P_p=\beta P_D$) it is clear that the term $\frac{\sigma^4M}{K\beta P_D}$ is also the dominating term in the denominator of the achievable rate expression in the no-phase-noise case. Therefore, the performance loss of both operation modes compared to the no-phase-noise scenario is small. The result is of particular importance since this work focuses mainly on the low SNR (per degree of freedom). This is also often the foreseen operating point of Massive MIMO \cite{CommMag13,YangGreen13}.

We proceed with a result on the sum-rate performance in the high-SNR regime.
\begin{myproposition}\label{prop:highSNR}
\emph{Saturation in the high-SNR regime.}
 In the presence of phase noise the effective information rate of the $k$-th user saturates for $\frac{P_D}{\sigma^2}\rightarrow\infty$ to the values, for synchronous operation 
\begin{align}\label{eq:highSNRs}
 R_k^s\rightarrow\frac{1}{N_c}\sum_{i\in\mathcal{I}_d}\log_2\left(1+\frac{M\alpha_k^2e^{-(\sigma_\phi^2+\sigma_\theta^2)(i-(k-1)L)}}{M\kappa_k[i]+\alpha_k\sum_{q=1}^{K}\alpha_q}\right),
\end{align}
and for non-synchronous operation 
\begin{align}\label{eq:highSNRns}
 R_k^{ns}\!\rightarrow\!\frac{1}{N_c}\!\sum_{i\in\mathcal{I}_d}\!\log_2\!\left(\!1\!+\!\frac{M\alpha_k^2e^{-(\sigma_\phi^2+\sigma_\theta^2)(i-(k-1)L)}}{M\alpha_k^2\varpi_k[i]+\xi_k[i]+\alpha_k\sum_{q=1}^{K}\alpha_q}\right).
\end{align}
\end{myproposition}
\begin{IEEEproof}
 The result follows from \eqref{eq:achRate} and the definitions of $R_k^s$ and $R_k^{ns}$ in \eqref{eq:userRateNS}.
\end{IEEEproof}
In the high-SNR regime, MRC is known to be suboptimal since intersymbol interference and multi-user interference dominate the effective noise term. Therefore saturation in the high-SNR regime is observed also in the no-phase-noise case due to the MRC reception strategy. 

A particularly desirable property of massive MIMO systems is the array power gain that they offer. 
The following proposition shows that 
 the phase-noise-impaired single-carrier massive MIMO uplink with TR-MRC receive processing and 
 estimated CSI offers an array gain of $O(\sqrt{M})$---the same scaling law as for flat fading channels
 without phase noise, derived in  \cite{Hien12TComm}.
\begin{myproposition}\label{prop:arrayGain} Under the assumptions made in Section~\ref{sec:sched}, an $O(\sqrt{M})$ array gain is achievable. 
\end{myproposition}
\begin{IEEEproof}
 We start by proving the proposition for the synchronous case. Let $P_D=\frac{E_u}{M^\eta}$, where $E_u$ is fixed. Based on the derived achievable rates in Proposition \ref{prop:AchRates}, we compute the maximum possible exponent, $\eta>0$, such that a fixed, non-zero rate for the $i$-th code of user $k$ can be achieved, while the transmit power of each user is scaled as $1/M^\eta$ with increasing $M$. From argument of the log expression in \eqref{eq:achRate}, i.e. the effective SINR, we have 
 \begin{align*}
&\text{SINR}_k[i]=\frac{\frac{E_uM\alpha_k^2}{\sigma^2 M^\eta}e^{-(\sigma_\phi^2+\sigma_\theta^2)(i-(k-1)L)}}{\frac{E_uM\kappa_k[i]}{\sigma^2 M^\eta}+\frac{E_u\alpha_k\sum_q\alpha_q}{\sigma^2 M^\eta}+\alpha_k+\frac{\sum_q\alpha_q}{\beta K}+\frac{M^\eta\sigma^2}{K\beta E_u}}\\
 &=\frac{\frac{E_u\alpha_k^2}{\sigma^2}e^{-(\sigma_\phi^2+\sigma_\theta^2)(i-(k-1)L)}}{\frac{E_u \kappa_k[i]}{\sigma^2}+\frac{E_u\alpha_k\sum_q\alpha_q }{M\sigma^2}+M^{\eta-1}\left(\alpha_k+\frac{\sum_q\alpha_q}{\beta K}\right)+\frac{M^{2\eta-1}\sigma^2}{K\beta E_u}}.
 \end{align*}
 As $M\rightarrow\infty$ we have $\lim_{M\rightarrow\infty}R_k^s[i]>0$ if $\eta-1\leq 0$ and $2\eta-1\leq 0\Rightarrow\eta\leq 1/2$. For $\eta=1/2$ the rate $R_k^s$ converges to the value (as $M\rightarrow\infty$) 
\begin{align}\label{eq:LargeMs}
 R_k^s\rightarrow\frac{1}{N_c}\sum_{i\in\mathcal{I}_d}\log_2\left(1+\frac{\frac{E_u}{\sigma^2}\alpha_k^2e^{-(\sigma_\phi^2+\sigma_\theta^2)(i-(k-1)L)}}{\frac{E_u}{\sigma^2}\kappa_k[i]+\frac{\sigma^2}{K\beta E_u}}\right).
\end{align}
Similarly, it can be proved that the array gain for the non-synchronous operation is $O(\sqrt{M})$ and the rate approaches (as $M\rightarrow\infty$) the value 
\begin{align}\label{eq:LargeMns}
 R_k^{ns}\rightarrow\frac{1}{N_c}\sum_{i\in\mathcal{I}_d}\log_2\left(1+\frac{\frac{E_u}{\sigma^2}\alpha_k^2e^{-(\sigma_\phi^2+\sigma_\theta^2)(i-(k-1)L)}}{\frac{E_u}{\sigma^2}\alpha_k^2\varpi_k[i]+\frac{\sigma^2}{K\beta E_u}}\right).
\end{align}
It is clear that for $\eta>1/2$ the achievable rates approach $0$ as $M\rightarrow\infty.$
\end{IEEEproof}

\section{Impact of Phase Noise Separately at the BS and at the User Terminals}\label{sec:SeparatePhN}

Based on the preceding analysis, we examine two special cases of particular interest. Namely, we study the impact on sum-rate performance, when there is phase noise only at the user terminals (UTs) and not at the BS (i.e. $\sigma_\phi^2=0$ and $\sigma_\theta^2\neq 0$) and vice versa  (i.e. $\sigma_\phi^2\neq 0$ and $\sigma_\theta^2=0$).

\subsection{Special Case 1: Phase Noise Only at the UTs, $\sigma_\phi^2=0$}\label{subsec:OnlyUTs}

If the oscillators at the BS are ideal,  there is no distinction between synchronous and non-synchronous operation. From \eqref{eq:achRate} it follows immediately that the lower bound in this case is given by
\begin{align}\label{eq:achRateOnlyUT}
R_k[i]&=\log_2\left(1+\frac{\frac{P_DM\alpha_k^2}{\sigma^2}e^{-\sigma_\theta^2(i-(k-1)L)}}{\frac{P_DM}{\sigma^2}\alpha_k^2\left(1-e^{-\sigma_\theta^2(i-(k-1)L)}\right)+\frac{C_k}{\sigma^2M}}\right).
\end{align}
In the high SNR limit the rate saturates at the value
\begin{align}\label{eq:highSNROnlyUT}
R_k[i]\!&\!\rightarrow\!\log_2\!\left(\!1\!+\!\frac{M\alpha_ke^{-\sigma_\theta^2(i-(k-1)L)}}{M\alpha_k\left(1-e^{-\sigma_\theta^2(i-(k-1)L)}\right)+\sum_{q=1}^{K}\alpha_q}\right).
\end{align}
Further, by scaling the transmit power as $P_D=E_u/\sqrt{M}$ we have the limiting expression as $M\rightarrow\infty$
\begin{align}\label{eq:largeMOnlyUT}
R_k[i]&\rightarrow\log_2\left(1+\frac{\frac{E_u}{\sigma^2}\alpha_k^2 e^{-\sigma_\theta^2(i-(k-1)L)}}{\frac{E_u}{\sigma^2}\alpha_k^2\left(1-e^{-\sigma_\theta^2(i-(k-1)L)}\right)+\frac{\sigma^2}{K\beta E_u}}\right).
\end{align}
In the following we provide an intuitive explanation of this similarity. Consider the link between user $k$ and the BS. Irrespectively of whether there is phase noise at the BS or not, the distortion in the received signal at each BS antenna due to the phase noise at the user adds up after TR-MRC processing, giving an additional interference term (see $\texttt{IF}_k[i]$ in \eqref{eq:effective_noise}) with a standard deviation that scales as $O(M)$.

\subsection{Special Case 2: Phase Noise Only at the BS, ($\sigma_\phi^2\neq 0$ and $\sigma_\theta^2=0$)}\label{subsec:OnlyBS}

In this case the achievable rate for the synchronous case is given by
\begin{align}\label{eq:achRateOnlyBSSynch}
R_k^s[i]=\log_2\left(1+\frac{\frac{P_DM\alpha_k^2}{\sigma^2}e^{-\sigma_\phi^2(i-(k-1)L)}}{\frac{P_DM}{\sigma^2}\xi_k[i]+\frac{C_k}{\sigma^2M}}\right),
\end{align}
and for the non-synchronous case
\begin{align}\label{eq:achRateOnlyBSNSynch}
R_k^{ns}[i]=\log_2\left(1+\frac{\frac{P_DM}{\sigma^2}\alpha_k^2e^{-\sigma_\phi^2(i-(k-1)L)}}{\frac{P_D}{\sigma^2}\xi_k[i]+\frac{C_k}{\sigma^2M}}\right).
\end{align}
In the high SNR regime the above rates saturate at the following values
\begin{align}\label{eq:highSNROnlyBSSynch}
R_k^s[i]&\rightarrow\log_2\left(1+\frac{M\alpha_k^2e^{-\sigma_\phi^2(i-(k-1)L)}}{M\xi_k[i]+\alpha_k\sum_{q=1}^{K}\alpha_q}\right)
\end{align}
\begin{align}\label{eq:highSNROnlyBSNSynch}
R_k^{ns}[i]&\rightarrow\log_2\left(1+\frac{M\alpha_k^2e^{-\sigma_\phi^2(i-(k-1)L)}}{\xi_k[i]+\alpha_k\sum_{q=1}^{K}\alpha_q}\right).
\end{align}
Further, by scaling the transmit power as $P_D=E_u/\sqrt{M}$ we have the limiting expressions as $M\rightarrow\infty$ for the synchronous operation
\begin{align}\label{eq:largeArrayOnlyBSSynch}
R^s_k[i]&\rightarrow\log_2\left(1+\frac{\frac{E_u}{\sigma^2}\alpha_k^2 e^{-\sigma_\phi^2(i-(k-1)L)}}{\frac{E_u}{\sigma^2}\xi_k[i]+\frac{\sigma^2}{K\beta E_u}}\right),
\end{align}
and for the non-synchronous operation
\begin{align}\label{eq:largeArrayOnlyBSNSynch}
R^{ns}_k[i]&\rightarrow\log_2\left(1+\left(\frac{E_u}{\sigma^2}\right)^2K\beta\alpha_k^2e^{-\sigma_\phi^2(i-(k-1)L)}\right).
\end{align}
The expressions in \eqref{eq:achRateOnlyBSSynch}, \eqref{eq:highSNROnlyBSSynch} and \eqref{eq:largeArrayOnlyBSSynch} are qualitatively similar to the case of phase noise only at the user terminals and with the general case with synchronous operation at the BS. In fact, it is the symmetric case as in Section \ref{subsec:OnlyUTs}. This behavior can be explained by arguments similar to the ones used there.

However, in the expressions for the non-synchronous operation \eqref{eq:achRateOnlyBSNSynch}, \eqref{eq:highSNROnlyBSNSynch} and \eqref{eq:largeArrayOnlyBSNSynch} we observe a fundamentally different behavior. Firstly, in \eqref{eq:highSNROnlyBSNSynch} we note that by increasing the number of BS antennas, we can increase the high-SNR saturation value of the achievable rate arbitrarily. In addition, from \eqref{eq:largeArrayOnlyBSNSynch} it is clear that in the large array regime we can arbitrarily increase the limiting expression by appropriately selecting the value $E_u$. These observations lead to the conclusion that the distortions introduced by independent oscillators at the BS asymptotically vanish, when TR-MRC reception is used. We remark that similar behavior was also noted in \cite{EmilHardware14}, where the authors demonstrate that the dominating impairment is the one at the hardware of the user equipment, while impairments at the BS from independent sources asymptotically vanish as $M\rightarrow\infty$.

\section{Numerical Examples}\label{sec:num_examples}

In this section, we present numerical examples of the main results presented in Sections \ref{sec:AchSumRate}--\ref{sec:SeparatePhN}. Throughout the section we selected $T_s=0.1\mu s$ and $f_c=2\text{ GHz}$, which correspond to typical values of wideband wireless communication systems, such as the WLAN IEEE 802.11. The reference value of the oscillator parameter $c_\phi$ (and $c_\theta$) is set to $c_\phi=4.7\times 10^{-18}(\text{rad Hz})^{-1}$, which also corresponds to a typical oscillator in WLAN IEEE 802.11 equipment \cite[Table 1]{Petrovic04}. However, we will refer to the standard deviation of the phase noise innovations, i.e. $\sigma_\phi$ and $\sigma_\theta$, since this is a more intuitive measure of the oscillator quality. For the parameters selected above and the relations in Section \ref{sub:PhNmodel}, $\sigma_\phi=0.49^o$. In typical cellular systems the delay spread is of the order of microseconds. We select $L=20$, which corresponds to $2 \mu s$ of delay spread for the selected symbol rate.  We selected the large scale fading factors as $\alpha_k=1,~\forall k\in\{1,...,K\}$, since the main purpose of this work is to understand the effect of phase noise and not of large scale fading. However, the same relations can be used with other choices of $\alpha_k$'s, when the study of particular propagation conditions is of interest. Further, we have selected a common power delay profile of every user as $d_{k,l}=e^{-0.35 l}/\sum_{p=0}^{L-1}e^{-0.35 p},~l=\{0,...,L-1\}$. We note that the power delay profile enters the rate expressions through the terms $\kappa_k[i]$ and $\xi_k[i]$ (see Proposition \ref{prop:varExpressions}). For most reasonable choices of $\sigma_\phi$ the choice of a particular PDP has a negligible effect on the achievable sum-rate. This choice of PDP and large scale fading is the same for all the figures that follow.

\begin{figure}
        \centering
        \begin{subfigure}[b]{0.5\textwidth}
                \includegraphics[width=\textwidth]{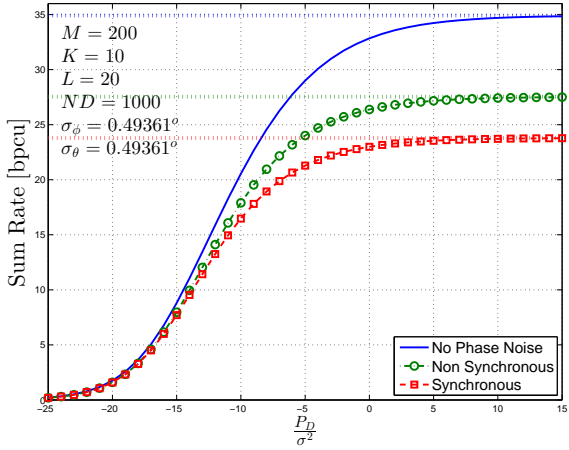}
                \caption{$\sigma_\phi=\sigma_\theta=0.49^o$}
                \label{fig:SRvsSNR1}
        \end{subfigure}
        ~ 
        \begin{subfigure}[b]{0.5\textwidth}
                \includegraphics[width=\textwidth]{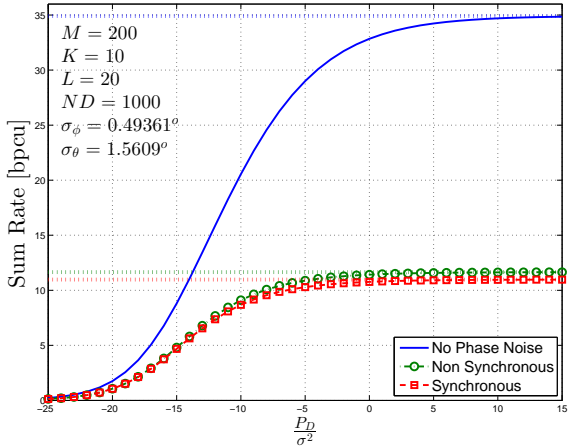}
                \caption{$\sigma_\phi=0.49^o$, $\sigma_\theta=1.56^o$}
                \label{fig:SRvsSNR2}
        \end{subfigure}
        ~ 
        \begin{subfigure}[b]{0.5\textwidth}
                \includegraphics[width=\textwidth]{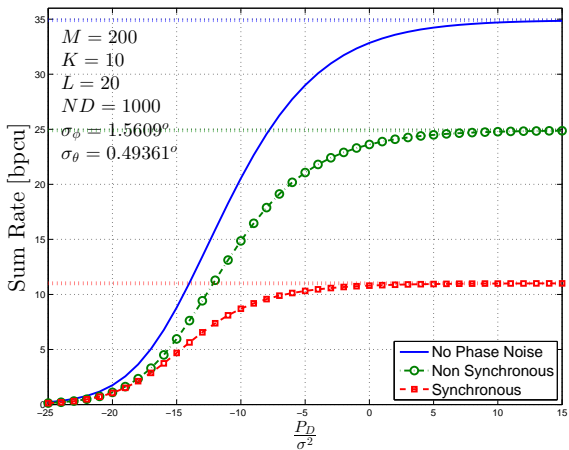}
                \caption{$\sigma_\phi=1.56^o$, $\sigma_\theta=0.49^o$}
                \label{fig:SRvsSNR3}
        \end{subfigure}
        \caption{Sum-rate as a function of $\frac{P_D}{\sigma^2}$ for $M=200$, $K=10$, $L=20$ and $N_D=1000$. The dotted vertical lines denote the high SNR asymptotic values of the achievable sum-rates.}\label{fig:SRvsSNR}
\end{figure}

In Fig. \ref{fig:SRvsSNR} the sum-rate performance of the system, as given by \eqref{eq:sumRateNS}, is plotted as a function of $\frac{P_D}{\sigma^2}$ for $N_D=1000$ with $M=200$, $K=10$. The sum-rate achieved without phase noise \eqref{eq:sumRateNophN} is plotted for the sake of comparison. We observe that at low SNR, the loss in sum-rate performance is insignificant. This observation supports our argument on the low SNR performance at the beginning of Section \ref{sec:mainResults}. We plot the sum-rate as a function of $\frac{P_D}{\sigma^2}$ for various choices of $\sigma_\phi$ and $\sigma_\theta$. It is clear from Fig. \ref{fig:SRvsSNR2} that when the phase noise at the user terminals is dominant both operation modes have similar performance. On the other hand, when the phase noise at the BS is dominant, as in Fig. \ref{fig:SRvsSNR3}, the sum-rate of the non-synchronous operation is significantly higher than the synchronous operation mode. This is in agreement with the discussion in Section \ref{sec:SeparatePhN}.

\begin{figure}
        \centering
        \includegraphics[width=0.5\textwidth]{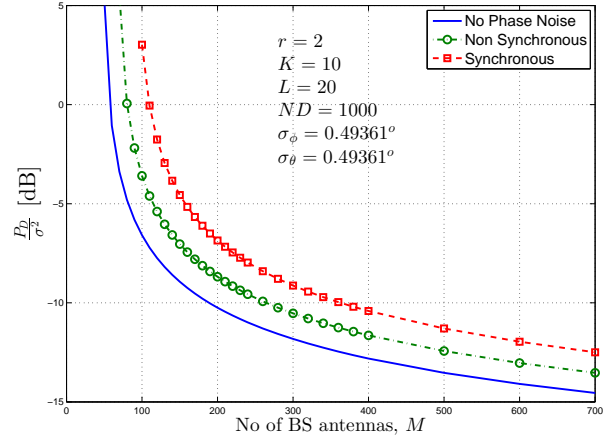}
        \caption{Minimum required $\frac{P_D}{\sigma^2}$ to achieve a fixed per-user information rate of $r = 2$ bpcu as a function of increasing $M$ for fixed $K=10$ users, $\sigma_\phi=\sigma_\theta=0.49^o$ and $N_D=1000$.}\label{fig:minSNRvsM}
\end{figure}

A significant desirable property of massive MIMO systems is the array power gain that they offer, facilitating the design of highly power-efficient communication systems \cite{Marzetta10, Hien12TComm, WCL12}. Proposition \ref{prop:arrayGain} extends this result to the case of single-carrier frequency-selective Massive MU-MIMO systems impaired with phase noise. The above observation is further supported through Fig. \ref{fig:minSNRvsM}, where the minimum per-user $\frac{P_D}{\sigma^2}$ required to achieve a fixed per-user information rate of $r = 2$ bpcu is plotted as a function of the number of BS antennas for $N_D = 1000$ and $K = 10$ for $\sigma_\phi=\sigma_\theta=0.49^o$. The plot for the phase-noise-free case is also given for the sake of comparison. We observe that by doubling the number of BS antennas we can reduce the per-user required $\frac{P_D}{\sigma^2}$ by 1.5dB, for sufficiently large $M$. This illustrates the validity of Proposition \ref{prop:arrayGain}.

From Fig. \ref{fig:minSNRvsM} we are motivated to study the gap in required $\frac{P_D}{\sigma^2}$ between the phase-noise-impaired cases and the no-phase-noise operation. In Table \ref{tab:PowerGap} we present numerical results on this gap. Each row corresponds to a different oscillator constant $c_\phi=c_\theta$, namely, $9.4\times 10^{-19}, ~4.7\times 10^{-18}\text{ and }2.35\times 10^{-17}(\text{rad Hz})^{-1}$, which correspond to standard deviation of phase noise innovations of $0.22^o$, $0.49^o$ and $1.1^o$, respectively. In order to give a more intuitive measure of the disturbance introduced by phase noise, we list the vertical $\frac{P_D}{\sigma^2}$ gap as a function of the standard deviation of the accumulated phase noise drift at a time difference of $N_D+L-1$ channel uses (i.e., the time difference between the end of the training phase and the end of the data phase). This result is shown in Table \ref{tab:PowerGap}. As expected, the performance gap is minimal for small phase noise drift and increases as the standard deviation of the phase noise drift increases.
\begin{table}[htbp]
\caption{Gap in required $\frac{P_D}{\sigma^2}$ due to phase noise for $N_D=1000$ and a fixed per-user information rate $r=1$ bpcu. The number of users is fixed to $K=10$.}
\begin{center}
\begin{tabular}{p{1.6cm}cccc}
\toprule
\multicolumn{ 5}{c}{Gap in required $\frac{P_D}{\sigma^2}$ [dB]} \\ \hline
{$\sigma_\phi\sqrt{N_D}$} & \multicolumn{ 2}{c}{Synchronous} & \multicolumn{ 2}{c}{Non-Synchronous}\\
{(degrees)} & M=500 & M=2500 & M=500 & M=2500 \\
\midrule
7.05\textdegree & 0.1174 & 0.1055 & 0.0828 & 0.0744 \\
15.76\textdegree & 0.6145 & 0.5492 & 0.4192 & 0.3753 \\
35.23\textdegree & 4.7459 & 3.9629 & 2.3071 & 2.0116 \\
\bottomrule
\end{tabular}
\end{center}
\label{tab:PowerGap}
\end{table}

It is also interesting to study the gap  in required $\frac{P_D}{\sigma^2}$ as a function of the desired per-user information rate. For this purpose we provide Table \ref{tab:PowerGapVarR}. There, we tabulate the gap  in required $\frac{P_D}{\sigma^2}$ in dB for various values of the per-user desired information rate for the synchronous and non-synchronous mode, for $N_D=1000$ channel uses, $\sigma_\phi=\sigma_\theta=0.49^o$, $K=10$ users and $M=500$ BS antennas. In the low spectral efficiency regime this gap is minimal. However, as the desired per-user information rate increases the gap increases at a faster rate. When the desired per-user information rate increases from 2 bpcu to 2.5 bpcu, which corresponds to 25\% increase, the gap in dB in the case of non-synchronous operation doubles, whereas in the synchronous operation mode the vertical gap increases more than two times. This happens because the desired per-user rate is close to the high-SNR saturation rate for the case of synchronous receivers\footnote{With the selected parameters, the high-SNR saturation value for the synchronous operation is 2.66 bpcu per user.}. As a result, a large increase in the transmit power is required in order to achieve the desired information rate.
\begin{table}[htbp]
\caption{Gap in required $\frac{P_D}{\sigma^2}$ due to phase noise for $N_D=1000$, $\sigma_\phi=\sigma_\theta=0.49^o$, $K=10$ users and $M=500$ BS antennas for various values of the desired per-user information rate in bits per channel use [bpcu].}
\begin{center}
\begin{tabular}{ccc}
\toprule
\multicolumn{ 3}{c}{Gap in required $\frac{P_D}{\sigma^2}$ [dB]} \\ \hline
Per-user rate & Synchronous & Non-Synchronous\\
\midrule
0.25 & 0.2768 & 0.2481 \\
0.5 & 0.3625 & 0.2941 \\
1 & 0.6145 & 0.4192 \\
2 & 2.2356 & 1.0987 \\
2.5 & 6.8694 & 2.1749 \\
\bottomrule
\end{tabular}
\end{center}
\label{tab:PowerGapVarR}
\end{table}

For fixed $M,~K$ and $L$ there is a fundamental trade-off between the length of the data interval, $N_D$, and the achievable sum-rate performance. A fraction $\frac{KL}{N_c}$ of each coherence interval is spent on training. Since a fixed time interval of $KL$ channel uses is required for channel estimation, a small data interval, $N_D$, leads to underutilization of the available resources, yielding a low sum-rate performance. As $N_D$ increases, more resources are utilized for the data transmission, increasing the sum-rate performance. However, as it can be seen from \eqref{eq:achRate}, $R_k^s[i]<R_k^s[i-1]$ and $R_k^{ns}[i]<R_k^{ns}[i-1]$, which implies that the gain of increasing the data interval diminishes with increasing $N_D$. In fact, the individual rates $R^s_k[i]$ and $R^{ns}_k[i]$ approach 0 as $i\rightarrow\infty$. This phenomenon occurs because with large $N_D$, the phase noise drift in the oscillators is so large such that there is a total loss of coherency between the received symbols during the data phase and the estimated channel at the beginning of the transmission block.

\begin{figure}
        \centering
        \includegraphics[width=0.5\textwidth]{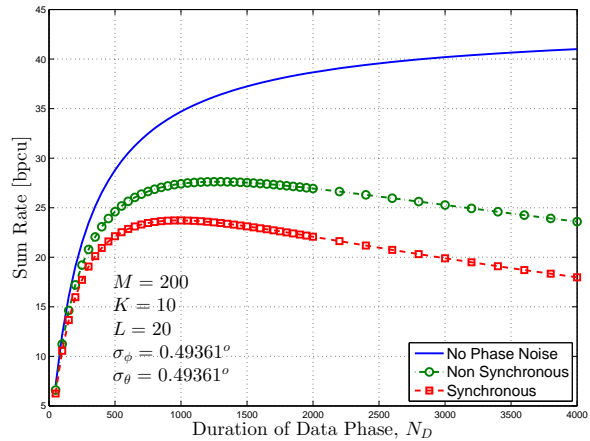}
        \caption{Sum-rate performance as a function of $N_D$, with fixed, $\sigma_\phi=\sigma_\theta=0.49^o$, $\frac{P_D}{\sigma^2}=10$ dB, $M=200$ BS antennas, $K=10$ users and $L=20$ taps.}\label{fig:SRvsND}
\end{figure}
In Fig. \ref{fig:SRvsND} the sum-rate performance is plotted as a function of $N_D$ for $\sigma_\phi=\sigma_\theta=0.49^o$. In the no-phase-noise case the optimal value of $N_D$ is infinity. However, there is a clear trade-off between the sum-rate and the length of the data interval in the phase-noise-impaired operation modes.

\begin{figure}
        \centering
        \includegraphics[width=0.5\textwidth]{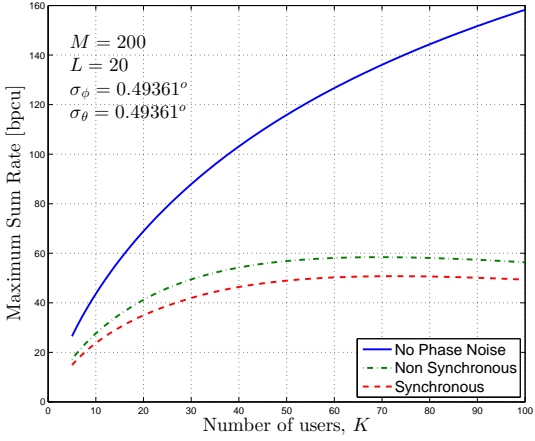}
        \caption{Maximum sum-rate performance as a function of $K$, with fixed $\frac{P_D}{\sigma^2}=10$ dB, $\sigma_\phi=\sigma_\theta=0.49^o$, $M=200$ BS antennas and $L=20$ taps. For each $K$, $N_D$ is optimally chosen.}\label{fig:maxSRvsK}
\end{figure}

Further insight can be obtained by considering the optimum number of scheduled users. In practice, the coherence interval is finite and therefore the training overhead upper-bounds the optimum number of scheduled users. Now, consider the case where the coherence interval is arbitrarily long. Then for the no-phase noise case, the optimal $N_D$ is unbounded. In that case one can increase the number of users, thereby achieving an increase in the sum-rate performance due to the spatial multiplexing of more users in the same time-frequency resource. In the presence of phase noise increasing the number of scheduled users, $K$, not only increases the length of the training overhead, but it also increases the phase drift between the estimated channel coefficients and the actual realizations of the effective channel impulse responses during the data interval. That is, by increasing the number of users, $K$, the partial loss of coherency between the estimated channel coefficients and the actual effective channels during data transmission is also increased. As a result, with increasing $K$ the increase in the achievable sum-rate during the data interval may eventually become insignificant to compensate for the reduction in sum-rate due to this partial loss of coherency. In Fig. \ref{fig:maxSRvsK}, for every $K$ the maximum achievable sum-rate performance is found by maximizing with respect to $N_D$ and, subsequently, this maximum sum-rate performance is plotted as a function of $K$ for $\frac{P_D}{\sigma^2}=10$ dB, $M=200$ BS antennas and $L=20$ taps for the no phase noise case, the synchronous operation mode and the non-synchronous operation mode. It is clear that the sum-rate performance is not monotonically increasing in the phase-noise-impaired cases as it is in the no phase noise case. However, it has a unimodal shape. This implies that in practice the optimum number of scheduled users is not only upper-bounded by the length of the coherence interval, but it is also upper-bounded as a consequence of the phase noise.

\section{Conclusions}
Phase noise is an inevitable hardware impairment in communication systems. We studied the effect of phase noise on the sum-rate performance of single-carrier transmission in a MU-MIMO uplink with an excess of BS antennas. Two distinct operation modes in terms of the phase noise processes at the BS antennas are considered, namely, synchronous and non-synchronous operation. Since the knowledge of the exact channel realizations is not available, CSI is acquired via uplink training. The BS uses TR-MRC receive processing to detect the information symbols. An analytical expression for the achievable sum-rate is rigorously derived for both operation modes. Based on the derived achievable sum-rates, we observe that it can be beneficial to use independent instead of fully synchronous phase noise sources. It is also shown that at low SNR, phase noise has little impact on the sum-rate performance. Further, the proposed receive processing achieves an $O(\sqrt{M})$ array power gain, extending earlier results where phase noise was not considered. Finally, due to the progressive phase noise drift in the oscillators, there is a fundamental trade-off between the length of the time interval used for data transmission and the sum-rate performance.

\appendix\label{appendix}
In this appendix we state the proof of Proposition \ref{prop:varExpressions}. For both operation modes, we have
\begin{align*}
&\texttt{Var}\left(\texttt{EN}_k[i]\right)\Define\mathbb{E}\left[|\texttt{EN}_k[i]-\mathbb{E}\left[\texttt{EN}_k[i]\right]|^2\right]=\texttt{Var}\left(\texttt{IF}_k[i]\right)\\
 &+\texttt{Var}\left(\texttt{ISI}_k [i]\right) + \texttt{Var}\left(\texttt{MUI}_k [i]\right)+\texttt{Var}\left(\texttt{AN}_k[i]\right)
\end{align*}
since the terms in $\texttt{EN}_k[i]$ are mutually uncorrelated. We start by computing the terms $\texttt{Var}\left(\texttt{ISI}_k [i]\right)$, $\texttt{Var}\left(\texttt{MUI}_k [i]\right)$, $\texttt{Var}\left(\texttt{AN}_k[i]\right)$ for the non-synchronous case, which are the same for both operation modes and conclude with the term $\texttt{Var}\left(\texttt{IF}_k[i]\right)$, the calculation of which is different depending on the operation mode. First we compute the variance of the ISI term.
\begin{align*}
 \mathbb{E}&[|\texttt{ISI}_k [i]|^2]=\mathbb{E}[|\sqrt{P_D}\sum_{m=1}^M\sum_{l=0}^{L-1}\sum_{\substack{q=0\\q\neq l}}^{L-1}g_{m,k,l}^*g_{m,k,q}\vartheta\!\left(\!\substack{m,k,k\\i,l,p}\!\right)\\
 &\cdot x_k[i+l-q]|^2]=P_D\sum_{m=1}^M\sum_{m'=1}^M\sum_{l=0}^{L-1}\sum_{l'=0}^{L-1}\sum_{\substack{p=0\\p\neq l}}^{L-1}\sum_{\substack{p'=0\\p'\neq l'}}^{L-1}\\
 &\cdot\mathbb{E}\left[g_{m,k,l}^*g_{m,k,p}g_{m',k,p'}^*g_{m',k,l'}\right]\\
 &\cdot\mathbb{E}\left[e^{-j(\phi_m[i+l]-\phi_{m'}[i+l']-\phi_m[(k-1)L+l]+\phi_{m'}[(k-1)L+l'])}\right]\\
 &\cdot\mathbb{E}\left[e^{j(\theta_k[i+l-p]-\theta_k[(k-1)L]-\theta_k[i+l'-p']+\theta_k[(k-1)L])}\right]\\ &\cdot\mathbb{E}\left[x_k[i+l-p]x_k^*[i+l'-p']\right]=P_D\sum_{m=1}^M\sum_{l=0}^{L-1}\sum_{\substack{q=0\\q\neq l}}^{L-1}d_{k,l}d_{k,q}\\
 &=P_DM\left(\alpha_k^2-\sum_{l=0}^{L-1}d_{k,l}^2\right), 
\end{align*}
where we have used the fact that the channel coefficients, the phase noise processes and the data symbols are mutually independent. The last step follows from the normalization of the PDP (see \eqref{eq:PDPnorm}). We will make use of these facts in all the following derivations as well. We proceed with the calculation of the multi-user interference.
\begin{align*}
 \mathbb{E}&[|\texttt{MUI}_k [i]|^2]=\mathbb{E}[|\sqrt{P_D}\!\sum_{m=1}^M\!\sum_{\substack{q=1\\q\neq k}}^K\!\sum_{l=0}^{L-1}\!\sum_{p=0}^{L-1}\!g_{m,k,l}^*g_{m,q,p}\vartheta\!\left(\!\substack{m,k,q\\i,l,p}\!\right)\\&\cdot x_q[i+l-p]|^2]=P_D\sum_{m=1}^M\sum_{m'=1}^M\sum_{\substack{q=1\\q\neq k}}^K\sum_{\substack{q'=1\\q'\neq k}}^K\sum_{l=0}^{L-1}\sum_{l'=0}^{L-1}\sum_{\substack{p=0\\p\neq l}}^{L-1}\sum_{\substack{p'=0\\p'\neq l}}^{L-1}\\
 &\cdot\mathbb{E}\left[g_{m,k,l}^*g_{m,q,p}g_{m',q',p'}^*g_{m',k,l'}\right]\\
 &\cdot\mathbb{E}\left[e^{-j(\phi_m[i+l]-\phi_{m'}[i+l']-\phi_m[(k-1)L+l]+\phi_{m'}[(k-1)L+l'])}\right]\\
 &\cdot\mathbb{E}\left[e^{j(\theta_q[i+l-p]-\theta_k[(k-1)L]-\theta_{q'}[i+l'-p']+\theta_k[(k-1)L])}\right]\\ &\cdot\mathbb{E}\left[x_q[i+l-p]x_{q'}^*[i+l'-p']\right]\\
 &=P_D\sum_{m=1}^M\sum_{\substack{q=1\\q\neq k}}^K\sum_{l=0}^{L-1}\sum_{p=0}^{L-1}d_{k,l}d_{q,p}=P_DM\alpha_k\sum_{\substack{q=1\\ q\neq k}}^{K}\alpha_q
\end{align*}
We conclude the first part of the proof with the calculation of the variance of the additive noise term.
\begin{align*}
 \mathbb{E}&[|\texttt{AN}_k[i]|^2] =\mathbb{E}[| \sqrt{\frac{P_D}{P_pKL}}\sum_{m=1}^M\sum_{q=1}^K\sum_{l=0}^{L-1}\sum_{p=0}^{L-1}g_{m,q,p}\\
 &\cdot e^{-j(\phi_m[i+l]-\theta_q[i+l-p])}n_m[(k-1)L+l]x_q[i+l-p]|^2]\\
 &+\mathbb{E}[|\sum_{m=1}^M\sum_{l=0}^{L-1}\hat g^*_{m,k,l}n_m[i+l]|^2]\\
 &=\frac{P_D}{P_pKL}\sum_{m=1}^M\sum_{m'=1}^M\sum_{q=1}^K\sum_{q'=1}^K\sum_{l=0}^{L-1}\sum_{l'=0}^{L-1}\sum_{p=0}^{L-1}\sum_{p'=0}^{L-1}
   \end{align*}
   \begin{align*}
 &\mathbb{E}[(g_{m,q,p}e^{-j(\phi_m[i+l]-\theta_q[i+l-p])}n_m[(\!k\!-\!1\!)L\!+\!l]x_q[\!i\!+\!l\!-\!p])\\
 &\cdot (g_{m',q',p'}e^{-j(\phi_{m'}[i+l']-\theta_{q'}[i+l'-p'])}n_{m'}[(k-1)L+l']\\
 &\cdot x_{q'}[i+l'-p'])^*]+\sigma^2\sum_{m=1}^M\sum_{l=0}^{L-1}\mathbb{E}[|\hat g_{m,k,l}|^2]\\
 &=\frac{P_D\sigma^2}{P_pKL}\sum_{m=1}^M\sum_{q=1}^K\sum_{l=0}^{L-1}\sum_{\substack{a=1-L\\0\leq l-a\leq L-1}}^{L-1}d_{q,l-a}\\
 &+\sigma^2\sum_{m=1}^M\sum_{l=0}^{L-1}\left(\frac{\sigma^2}{P_pKL}+\mathbb{E}[|g_{m,k,l}|^2]\right)\\ 
 &=\sigma^2M\left(\frac{P_D}{P_pK}\sum_{q=1}^{K}\alpha_q+\frac{\sigma^2}{P_pK}+\alpha_k\right) 
\end{align*}
We proceed by calculating the variance of the term $\texttt{IF}_k[i]$. It holds
\begin{align*}
\texttt{Var}(\texttt{IF}_k[i]) &= \mathbb{E}\left[|(A_k[i]-\mathbb{E}[A_k[i]])x_k[i]|^2\right]\\
&=\mathbb{E}\left[\left|A_k[i]\right|^2\right]-\left|\mathbb{E}\left[A_k[i]\right]\right|^2.
\end{align*}
Based on the result of Proposition \ref{prop:meanExpressions} it is sufficient to calculate $\mathbb{E}\left[\left|A_k[i]\right|^2\right]$ for each operation mode. We start with the synchronous operation.

\begin{align*}
\mathbb{E}&\left[\left|A_k[i]\right|^2\right]=P_D\sum_{m=1}^M\sum_{l=0}^{L-1}\mathbb{E}[|g_{m,k,l}|^4]\\
&+P_D\sum_{m=1}^M\sum_{l=0}^{L-1}\sum_{\substack{l'=0\\l'\neq l}}^{L-1}\mathbb{E}[|g_{m,k,l}|^2]\mathbb{E}[|g_{m,k,l'}|^2]\\&\cdot\mathbb{E}[e^{-j(\phi[i+l]-\phi[i+l']-\phi[(k-1)L+l]+\phi[(k-1)L+l'])}]\\&+ P_D\sum_{m=1}^M\sum_{\substack{m'=1\\m'\neq m}}^M\sum_{l=0}^{L-1}\sum_{l'=0}^{L-1}\mathbb{E}[|g_{m,k,l}|^2]\mathbb{E}[|g_{m',k,l'}|^2]\\
 &\cdot\mathbb{E}[e^{-j(\phi[i+l]-\phi[i+l']-\phi[(k-1)L+l]+\phi[(k-1)L+l'])}]\\
 &=P_DM\sum_{l=0}^{L-1}2d^2_{k,l}+P_DM\sum_{l=0}^{L-1}\sum_{\substack{l'=0\\l'\neq l}}^{L-1}d_{k,l}d_{k,l'}e^{-\sigma_\phi^2|l-l'|}\\
 &+P_DM(M-1)\sum_{l=0}^{L-1}\sum_{l'=0}^{L-1}d_{k,l}d_{k,l'}e^{-\sigma_\phi^2|l-l'|}\\
  &=P_DM\sum_{l=0}^{L-1}d^2_{k,l}+P_DM^2\sum_{l=0}^{L-1}\sum_{l'=0}^{L-1}d_{k,l}d_{k,l'}e^{-\sigma_\phi^2|l-l'|}
  \end{align*}
  
Finally, for the synchronous operation, the effective noise variance, is given by

\begin{align*}
 \varsigma_k^s[i]\Define\texttt{Var}(\texttt{EN}^s_k[i])&= P_DM^2\kappa_k[i]+C_k.
\end{align*}
We conclude with the calculation of the term $\mathbb{E}\left[\left|A_k[i]\right|^2\right]$ for the non-synchronous mode.
\begin{align*}
\mathbb{E}&\left[\left|A_k[i]\right|^2\right]=P_D\sum_{m=1}^M\sum_{l=0}^{L-1}\mathbb{E}[|g_{m,k,l}|^4]\\
&+P_D\sum_{m=1}^M\sum_{l=0}^{L-1}\sum_{\substack{l'=0\\l'\neq l}}^{L-1}\mathbb{E}[|g_{m,k,l}|^2]\mathbb{E}[|g_{m,k,l'}|^2]\\
&\cdot\mathbb{E}[e^{-j(\phi_m[i+l]-\phi_{m}[i+l']-\phi_m[(k-1)L+l]+\phi_{m}[(k-1)L+l'])}]\\
&+ P_D\sum_{m=1}^M\sum_{\substack{m'=1\\m'\neq m}}^M\sum_{l=0}^{L-1}\sum_{l'=0}^{L-1}\mathbb{E}[|g_{m,k,l}|^2]\mathbb{E}[|g_{m',k,l'}|^2]\\
 &\cdot\mathbb{E}[e^{-j(\phi_m[i+l]-\phi_{m'}[i+l']-\phi_m[(k-1)L+l]+\phi_{m'}[(k-1)L+l'])}]\\
 &=P_DM\sum_{l=0}^{L-1}2d^2_{k,l}+P_DM\sum_{l=0}^{L-1}\sum_{\substack{l'=0\\l'\neq l}}^{L-1}d_{k,l}d_{k,l'}e^{-\sigma_\phi^2|l-l'|}\\
  &+P_DM(M-1)\sum_{l=0}^{L-1}\sum_{l'=0}^{L-1}d_{k,l}d_{k,l'}e^{-\sigma_\phi^2(i-(k-1)L)}\\
  &=P_DM\sum_{l=0}^{L-1}d^2_{k,l}+P_DM\sum_{l=0}^{L-1}\sum_{l'=0}^{L-1}d_{k,l}d_{k,l'}e^{-\sigma_\phi^2|l-l'|}\\
  &+P_DM(M-1)\alpha_k^2e^{-\sigma_\phi^2(i-(k-1)L)}.
\end{align*}

The variance for the non-synchronous operation is

\begin{align*}
  \varsigma_k^{ns}[i]\Define\texttt{Var}(\texttt{EN}^{ns}_k[i])= P_DM\xi_k[i]
 +P_DM^2\varpi_k[i]+C_k.
\end{align*}
\bibliographystyle{ieeetr}
\bibliography{phNbib}

\end{document}